\documentclass[apj,iop]{emulateapj}

\newcommand{\kev}{keV}
\newcommand{\fe}{Fe~K$\alpha$}
\newcommand{\etal}{et al.}
\newcommand{\threec}{3C~382}

\newcommand{\suzaku}{\textit{Suzaku}}
\newcommand{\chandra}{\textit{Chandra}}
\newcommand{\xmm}{\textit{XMM-Newton}}
\newcommand{\nustar}{\textit{NuSTAR}}
\newcommand{\swift}{\textit{Swift}}

\shorttitle{The Comptonizing Corona of 3C~382}
\shortauthors{Ballantyne \etal}

\usepackage{times}

\begin{document} 

\title{\textit{NuSTAR} Reveals the Comptonizing Corona of the
  Broad-Line Radio Galaxy 3C~382}


\author{D. R. Ballantyne\altaffilmark{1},
  J. M. Bollenbacher\altaffilmark{1}, L. W. Brenneman\altaffilmark{2},
K. K. Madsen\altaffilmark{3}, M. Balokovi\'{c}\altaffilmark{3},
S. E. Boggs\altaffilmark{4}, F. E. Christensen\altaffilmark{5},
W. W. Craig\altaffilmark{5,6}, P. Gandhi\altaffilmark{7}, C. J. Hailey\altaffilmark{8},
F. A. Harrison\altaffilmark{3}, A. M. Lohfink\altaffilmark{9},
A. Marinucci\altaffilmark{10},  C. B. Markwardt\altaffilmark{11},
D. Stern\altaffilmark{12}, D. J. Walton\altaffilmark{3}, and
W. W. Zhang\altaffilmark{11}}
\altaffiltext{1}{Center for Relativistic Astrophysics, School of Physics,
  Georgia Institute of Technology, Atlanta, GA 30332;
  david.ballantyne@physics.gatech.edu}
\altaffiltext{2}{Harvard-Smithsonian CfA, 60 Garden St. MS-67,
  Cambridge, MA 02138, USA}
\altaffiltext{3}{Cahill Center for Astronomy and Astrophysics,
  California Institute of Technology, Pasadena, CA 91125, USA}
\altaffiltext{4}{Space Science Laboratory, University of California,
  Berkeley, California 94720, USA}
\altaffiltext{5}{DTU SpaceNational Space Institute, Technical
  University of Denmark, Elektrovej 327, 2800 Lyngby, Denmark}
\altaffiltext{6}{Lawrence Livermore National Laboratory, Livermore,
  California 94550, USA}
\altaffiltext{7}{Department of Physics, University of Durham, South
  Road, Durham DH1 3LE, UK}
\altaffiltext{8}{Columbia Astrophysics Laboratory, Columbia
  University, New York, New York 10027, USA}
\altaffiltext{9}{Department of Astronomy, University of Maryland,
  College Park, MD 20742-2421 USA}
\altaffiltext{10}{Dipartimento di Matematica e Fisica, Universit\`{a}
  degli Studi Roma Tre, via della Vasca Navale 84, 00146 Roma, Italy}
\altaffiltext{11}{NASA Goddard Space Flight Center, Greenbelt,
  Maryland 20771, USA}
\altaffiltext{12}{Jet Propulsion Laboratory, California Institute of
  Technology, Pasadena, CA 91109, USA}

\begin{abstract}
Broad-line radio galaxies (BLRGs) are active galactic nuclei that produce
powerful, large-scale radio jets, but appear as Seyfert~1 galaxies in
their optical spectra. In the X-ray band, BLRGs also appear like
Seyfert galaxies, but with flatter spectra and weaker reflection features. One explanation for these properties is that the X-ray continuum is
diluted by emission from the jet. Here, we present two
\nustar\ observations of the BLRG \threec\ that show clear evidence
that the continuum of this source is dominated by thermal
Comptonization, as in Seyfert~1 galaxies. The two observations were
separated by over a year and found
\threec\ in different states separated by a factor of $1.7$ in flux. The
lower flux spectrum has a photon-index of $\Gamma=1.68^{+0.03}_{-0.02}$,
while the photon-index of the higher flux spectrum is
$\Gamma=1.78^{+0.02}_{-0.03}$. Thermal and anisotropic Comptonization models provide an excellent fit to both spectra and show that the
coronal plasma cooled from $kT_e=330\pm 30$~keV in the low flux data to
$231^{+50}_{-88}$~keV in the high flux observation. This cooling behavior is typical of Comptonizing
corona in Seyfert galaxies and is
distinct from the variations observed in jet-dominated
sources. In the high flux observation, simultaneous \swift\ data are leveraged to obtain a broadband spectral energy distribution and
indicates that the corona intercepts $\sim 10$\% of the optical and ultraviolet emitting accretion disk. \threec\ exhibits very weak
reflection features, with no detectable relativistic \fe\ line, that
may be best explained by an outflowing corona combined with an ionized
inner accretion disk.
\end{abstract}


\section{Introduction}
\label{sect:intro}
An important problem in extragalactic astrophysics is
understanding the physical triggers that allow a small fraction of
active galactic nuclei (AGN) to produce powerful, large-scale radio
jets. X-ray spectroscopic observations of AGNs are
the most direct probe of the complex interactions of magnetic fields,
fluid dynamics and relativistic physics that occur close to the
central black hole \citep[e.g.,][]{rn03}, and thus have the potential to make
significant progress in elucidating the physical triggers of jetted
AGNs \citep[e.g.,][]{ball07,loh13}. Observations of the brightest unobscured jetted AGNs, the
broad-line radio galaxies (BLRGs), have consistently shown weaker
reflection features and flatter X-ray spectra than typical Seyfert 1 galaxies
\citep[e.g.,][]{sem99,esm00,grandi01,zg01,ball07,sam09,evans10}. In
particular, the \fe\ lines are often observed to be narrow, low
equivalent width (EW) features with a relativistic
component from the inner accretion disk only rarely detectable \citep[e.g.,][]{loh13}. A number of explanations have
been proposed to account for the weak reflection signatures in BLRGs
including high inner disk ionization \citep{brf02}, a change in the inner disk
geometry \citep{esm00,loh13}, obscuration of the central
accretion flow by the jet \citep{sam09}, black holes with retrograde
spin \citep{ges10,evans10}, and dilution of the X-ray spectrum
by jet emission \citep{gum02}.

The broadband ($3$--$79$~keV), high sensitivity spectra provided by
the focusing hard X-ray telescopes on the
\textit{Nuclear Spectroscopic Telescope Array} (\nustar;
\citealt{harr13}) observatory have the potential to be
crucial in determining the correct interpretation of the BLRG
spectra. The wide energy range allows for an accurate separation of the
reflection and primary continua, as well as a more precise
determination of the ionization state of the reflector
\citep[e.g.,][]{ris13,bren14,bren14b,mar14}. Moreover, if the X-ray continuum of a BLRG is dominated by a
Comptonizing corona (as in Seyfert 1 galaxies; e.g.,
\citealt{pet00,pet01}) then the high-energy sensitivity of
\nustar\ will allow for measurements of the temperature ($kT_e$) and optical depth
($\tau$) of the plasma. In this way, \nustar\ data may finally start
placing some physical constraints on the interpretation of the X-ray
spectra of BLRGs.

Here, we report on two \nustar\ observations of the bright
($F_{2-10\ \mathrm{keV}} \approx 3$--$6\times
10^{-11}$~erg~cm$^{-2}$~s$^{-1}$; \citealt{glio07}),
nearby ($z=0.058$) BLRG \threec. The source was detected in two
different flux states that were separated by a factor of $1.7$, thus
spanning the historical observed range. These observations allow detailed Comptonization modeling for
different radiative conditions within the object. The next section describes the details
of the \nustar\ observations, and the spectral analysis is presented
in Sect.~\ref{sect:analyze}. Finally, Sect.~\ref{sect:discuss}
contains a discussion of the results.

\section{Observations}
\label{sect:obs}
\begin{figure*}
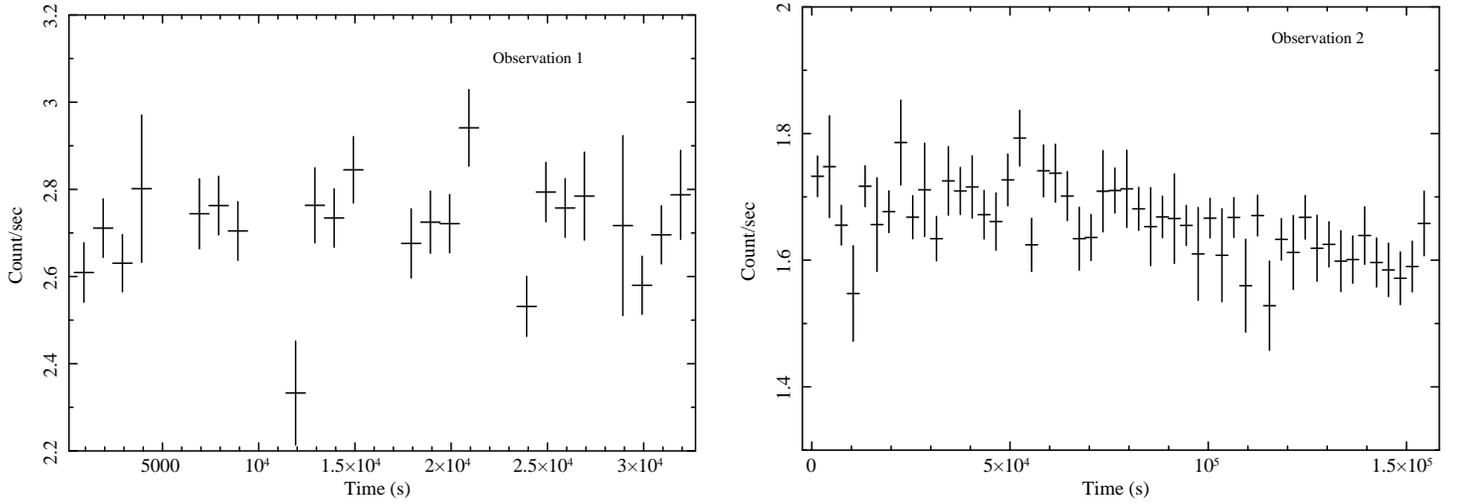

\centerline{
\includegraphics[width=0.375\textwidth,angle=-90]{obs1lc.ps}
\includegraphics[width=0.375\textwidth,angle=-90]{revisedobs2lc.ps}
}
\caption{Fullband ($3$--$79$~keV), background-subtracted
  \nustar\ lightcurves of \threec\ during
  Observation 1 (Left; 1000~s binning) and 2 (Right; 3000~s binning). Data from FPMA and FPMB were
  summed to determine both lightcurves. The lightcurves are well
  described by a model with a constant count-rate:
  $\chi^2/$dof$=24/22$ and $\chi^2/$dof$=51/51$ for Observation 1 and 2,
  respectively. The low point in the light curve of Observation 1 only
appears in the FPMA data, so does not indicate true variability from \threec.}
\label{fig:lc}
\end{figure*}

\begin{deluxetable*}{lccccc}
\tablewidth{0pt}
\tablecaption{\label{table:log} 3C~382 Observation Log}
\tablecolumns{6}
\tablehead{
\colhead{\phantom{i}} & \colhead{Telescope} & \colhead{Observation ID}
& \colhead{UT Start Date} & \colhead{Exposure (ks)} & \colhead{Counts}}
\startdata
Observation 1 & \nustar\ & 60061286002 & 2012 Sept 18 & 16.6/16.6 &
20367/19988 \\
 (High flux) & \swift-XRT & 00080217001 & 2012 Sept 18 & 6.6 & 3678 \\[0.25cm]
Observation 2 & \nustar\ & 60001084002 & 2013 Dec 18 & 82.6/82.4 &
63596/61219 \\
(Low flux) & & & & &
\enddata
\tablecomments{For the \nustar\ observations, exposure times and
  counts are listed for the two focal plane modules (FPMA/FPMB). The
  \nustar\ counts are the $3$--$79$~\kev\ background subtracted
  counts. Similarly, the background subtracted $0.3$--$7$~keV counts
  are indicated for the \swift-XRT data. }
\end{deluxetable*}
Table~\ref{table:log} lists the details of the \nustar\ observations of \threec. Both \nustar\ datasets were reduced and spectral products extracted
following standard procedures and using NuSTARDAS v.1.3.1 (included
in HEASoft v.6.15.1) and \nustar\ CALDB 20131223. Source data from
both focal plane modules (FPMA and FPMB) were extracted using a circular region of radius $60$\arcsec\ centered on \threec. A
$80$\arcsec\ radius source-free region on the same chip of the
detector plane was used to extract the background in Observation 1,
while a $95$\arcsec\ radius region was used for the background region
in Observation 2. The fullband, background-subtracted lightcurves of
\threec\ (summed from FPMA and FPMB) are shown in
Figure~\ref{fig:lc}.
\threec\ exhibited very little variability during
both observations, consistent with earlier observations of this
source \citep{grandi01,glio07,sam11}. Fitting the light-curves (with 1000~s
binning for Observation 1, and 3000~s binning for Observation 2) with a constant
gives a $\chi^2$ per degree of freedom (dof) of $\chi^2/$dof$=24/22$
and $\chi^2/$dof$=51/51$ during Observation 1 and 2, respectively. The
lack of variability during both observations allows for time-averaged
spectra to be accumulated for analysis. 

As the Comptonization modeling requires an accurate spectral
shape at high energies, all \nustar\ spectra were re-binned to have a
minimum signal-to-noise ratio of 5 in each spectral bin. For
Observation 1, this limits the energy range of the spectra to
$3$--$65$~keV (FPMA) and $3$--$63$~keV (FPMB). The data from the
longer Observation 2 is limited to the energy range of
$3$--$68$~\kev\ (FPMA) and $3$--$65$~\kev\ (FPMB).

\threec\ is one of the targets of \nustar's serendipitous survey
program \citep{alex13,harr13,balo14} that is utilizing bright
\swift-BAT sources as a means to perform a shallow, wide-field
survey. As part of this program, \swift-XRT performed a 6.6~ks simultaneous observation
of \threec\ during Observation 1. The \swift-XRT data were reduced and
a time-averaged spectrum extracted by automatic analysis tools
produced by the \swift-XRT team \citep{evans09}. The response matrix used
for the analysis was \texttt{swxpc0to12s6\_20010101v013.rmf}. To
facilitate $\chi^2$ fitting, the \swift-XRT spectrum was grouped to a
minimum of 25 counts/bin, and, due to a lack of counts at high
energies, only data between $0.3$ and $7$~\kev\ are included in the
analysis. 

The \swift-UV/Optical Telescope (UVOT) was also operating
during the observation and cycled through the V, B, U, UW1, UM2 and UW2 filters, taking one
image for each of the filters. The UVOT data were downloaded from the
HEASARC website and the \texttt{uvotsource} tool was used
to perform aperture photometry. The tool returns the coincidence
loss-corrected fluxes for a given source and background region. Here,
the source region was selected to be circular with a 5.0 arcsec radius and centered on \threec. The
background flux was obtained from a source-free circular region with a
20 arcsec radius in the vicinity of the source. The optical/UV fluxes
were corrected for Galactic extinction using the reddening laws of
\citet{cardelli:89a} and \citet{odonnell:94a} with a
$E(B-V)=0.0598$ \citep{schlegel:98a}. Assuming $R_V=3.16$ and the same
reddening law, the fluxes were then corrected for the internal
extinction of $E(B-V) \approx 0.23$ \citep{tadhunter:86a}. The
resulting fluxes are used to analyze the multi-wavelength spectral
energy distribution (SED) of \threec\ during Observation 1 (Sect.~\ref{sub:coronal}). There is
likely a small contribution from the host galaxy in the derived
fluxes, however it is clear that the host galaxy of \threec\ is weak
at UV wavelengths and the flux is dominated by the
nucleus \citep{allen:02a}, thus we omit any host correction for the
purpose of the simple comparison with the predicted SEDs.

XSPEC v.12.8.1l \citep{arn96} is used for all X-ray spectral fitting. Spectra from
FPMA and FPMB are fit simultaneously for both \nustar\ observations,
with a normalization constant left free to account for the slight calibration
differences between the two modules. The \swift-XRT spectrum is
included in the analysis of Observation 1, with an additional normalization
constant left free to vary in the fits. $\chi^2$ statistics are used to determine the best model
description of the data, and a $\Delta \chi^2=2.71$ criterion (i.e., a
90\% confidence range for one parameter of interest) is used to to
determine the error-bars.  The
following cosmological parameters are assumed: $\Omega_M=0.27$,
$\Omega_{\Lambda}=0.73$ and $H_{0}=70$~km~s$^{-1}$~Mpc$^{-1}$. 

\section{Spectral Analysis}
\label{sect:analyze}
Galactic absorption with a column density of
$N_{\mathrm{H}}=6.98\times 10^{20}$~cm$^{-2}$ \citep{kalb05} is included in
all spectral models using the \texttt{TBabs} model \citep{wilms00}. In addition,
\threec\ is observed to have a weak, highly ionized warm absorber
with $N_{\mathrm{H}} \approx 1.4\times 10^{21}$~cm$^{-2}$ and $\log
\xi=2.5$, where $\xi$ is the ionization parameter of the absorbing gas
\citep{torr10,sam11,walton13}. Although the effects of the warm absorber are minor in the
\nustar\ band, it is included in the spectral model (with the above
parameters) using a grid of XSTAR models calculated by \citet{walton13}. High
resolution \chandra\ observations of \threec\ indicate a diffuse halo
of emission peaking at a radius of $\approx 10$--$20$\arcsec\ from the point source
\citep{glio07}. The temperature and luminosity of this emission is too low to
be important in the \nustar\ energy band, nor can it account for the
observed soft excess \citep{torr10}, so it is omitted from the spectral
modeling.

\subsection{Observation 2 --- Low Flux}
\label{sub:obs1}
\begin{figure*}
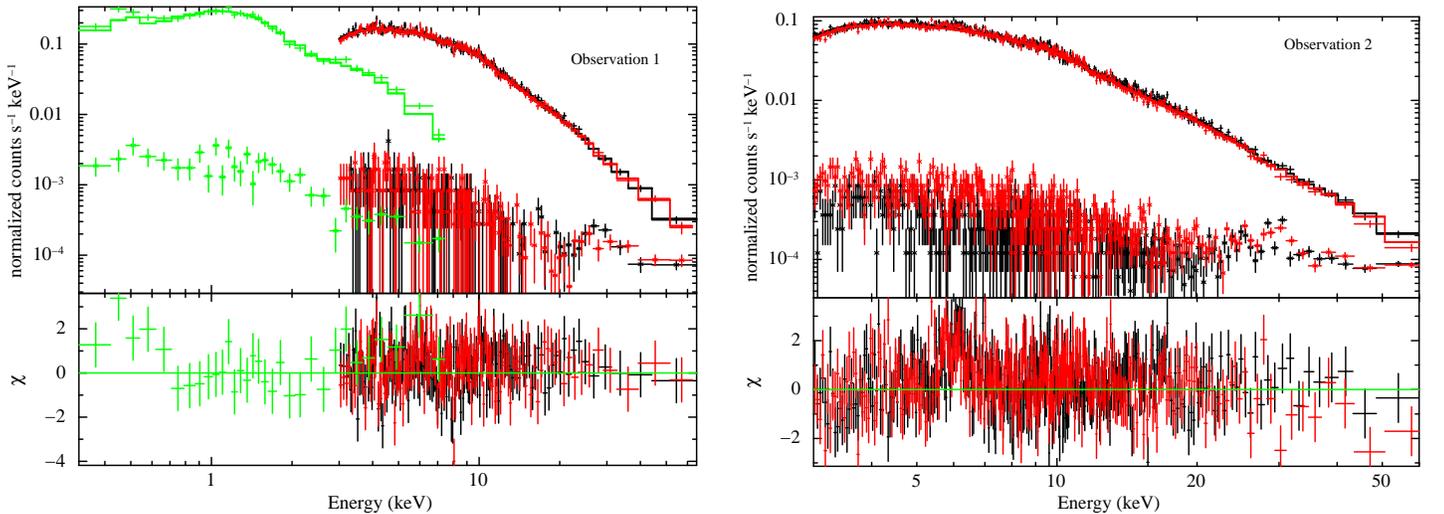

\centerline{
\includegraphics[width=0.375\textwidth,angle=-90]{newobs1.ps}
\includegraphics[width=0.375\textwidth,angle=-90]{newobs2.ps}
}
\caption{(Left) The top panel plots the count rate spectrum
  of \threec\ and the background obtained from Observation 1 for the \nustar\ FPMA (black) and FPMB (red)
  detectors. The \swift-XRT data are
  shown in green. The lower panel shows the residuals (in units of standard
  deviations) when the spectra are fit with a power-law modified
  by Galactic absorption and a weak warm absorber. The data between
  $4$ and $7.5$~\kev\ were not included in the fit. (Right) Similar to
  the left-hand panel, but for Observation 2. }
\label{fig:data}
\end{figure*}

\begin{figure}
\includegraphics[width=0.324\textwidth,angle=-90]{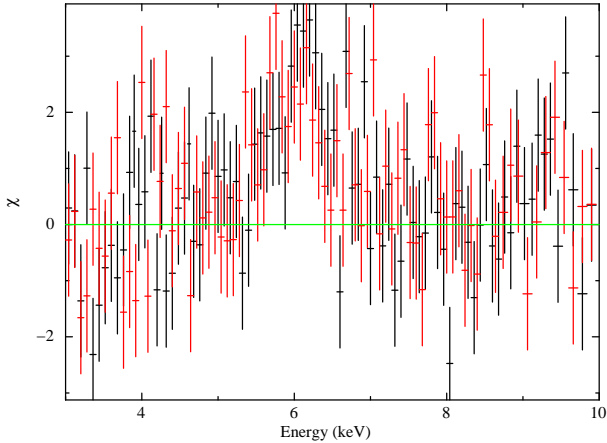}
\caption{As in Figure~\ref{fig:data} (Right), but zooming in and
  rebinning to emphasize the residuals in the \fe\ band. }
\label{fig:fek}
\end{figure}

\begin{figure}
\includegraphics[width=0.5\textwidth]{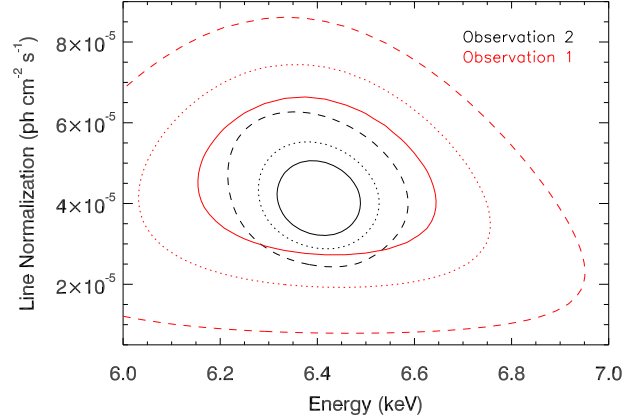}
\caption{The 68\% (solid), 90\% (dotted) and 95\% (dashed) joint confidence
  contours on the \fe\ line flux and energy calculated for
  Observations 1 (red) and 2 (black) using the cutoff power-law plus
  Gaussian line model. Both observations measure a similar line flux
  and energy. The line normalization is also consistent with many
  earlier measurements from \threec\ (Sect.~\ref{sub:prior}).}
\label{fig:linecontours}
\end{figure}

We begin by analyzing the \nustar\ spectra from Observation~2 as it
has the largest number of counts. The right-hand panel of
Fig.~\ref{fig:data} presents the count spectra and the residuals to a simple
power-law fit when the $4$--$7.5$~\kev\ data are ignored.
The fit is poor ($\chi^2/$dof$=1126/991$),
with a clear residual at the energy of the \fe\ line
(Figure~\ref{fig:fek}) and signs of a
turn down at high energies.
Replacing the power-law with a cutoff
power-law improves the fit to $\chi^2/$dof$=1108/990$ (significant at
$>99.99$\% according to the F-test) with
$E_{\mathrm{cut}}=202^{+132}_{-58}$~\kev. The addition of a narrow
Gaussian \fe\ line, fixing its width to $\sigma=90$~eV \citep{glio07}
and energy to $6.4$~\kev, further improves the fit
($\chi^2/$dof$=995/989$; significant at $\gg 99.99$\% according
to the F-test). While allowing the energy of the line to float does
not significantly improve the fit, thawing the width of the line
results in dropping the $\chi^2$ to $\chi^2/$dof$=980/988$ with
$\sigma=0.34^{+0.15}_{-0.12}$~\kev\ (F-test probability $>
99.9$\%). The cutoff energy increases to
$E_{\mathrm{cut}}=214^{+147}_{-63}$~\kev. At this stage, no significant residuals are
observed in the data. Re-introducing the narrow ($\sigma=90$~eV)
\fe\ line to the model with the broadened line 
does not result in any improvement to the fit. The best fit parameters from the simple cutoff
power-law plus Gaussian model are shown in Table~\ref{table:fits}, and
contours of line-flux versus line energy are shown in
Figure~\ref{fig:linecontours}. 
\begin{deluxetable}{lcc}
\tablewidth{0.5\textwidth}
\tablecaption{\label{table:fits} 3C~382 Spectral Fitting Results}
\tablecolumns{3}
\tablehead{
\colhead{\phantom{i}} & \colhead{Observation 1} & \colhead{Observation 2}}
\startdata
\cutinhead{Cutoff power-law + $6.4$~\kev\ Gaussian}
$\Gamma$ & $1.78^{+0.02}_{-0.03}$ & $1.68^{+0.03}_{-0.02}$ \\
$E_{\mathrm{cut}}$ (keV) & $> 190$ & $214^{+147}_{-63}$ \\
$\sigma$ (keV) & $0.37^{+0.3}_{-0.2}$ & $0.34^{+0.15}_{-0.12}$ \\
\fe\ flux (ph cm$^{-2}$ s$^{-1}$) & $4.8^{+2.5}_{-2.4}\times 10^{-5}$
& $4.1^{+1.1}_{-0.9}\times 10^{-5}$ \\
EW (eV) & $79^{+52}_{-44}$ & $114^{+30}_{-26}$ \\
$\chi^2/$dof & $780/786$ & $980/988$ \\
\cutinhead{\texttt{compps} + $6.4$~\kev\ Gaussian}
$kT_e$ (keV) & $231^{+50}_{-88}$ & $330\pm 30$ \\
$y$ & $0.40^{+0.07}_{-0.06}$ & $0.38^{+0.04}_{-0.05}$ \\
$\tau$ & $0.23$ & $0.15$ \\
$\sigma$ (keV) & $0.39^{+0.27}_{-0.25}$ & $0.29^{+0.14}_{-0.11}$\\
\fe\ flux (ph cm$^{-2}$ s$^{-1}$) & $4.5^{+2.6}_{-2.4}\times 10^{-5}$
& $3.5^{+1.0}_{-0.8}\times 10^{-5}$ \\
EW (eV) & $75^{+55}_{-52}$ & $98\pm 25$ \\
$\chi^2/$dof & $778/786$ & $976/988$ \\
$F_{2-10\ \mathrm{keV}}$ (erg cm$^{-2}$~s$^{-1}$)& $5\times 10^{-11}$ & $2.9\times 10^{-11}$ \\
\enddata
\tablecomments{A Galactic absorber with $N_{\mathrm{H}} = 6.98\times
  10^{20}$~cm$^{-2}$ \citep{kalb05} is included in all fits using the \texttt{TBabs}
  model \citep{wilms00}. Similarly, a weak warm absorber with
  $N_{\mathrm{H}}=1.4\times 10^{21}$~cm$^{-2}$ and $\log \xi =2.5$
  \citep[e.g.,][]{torr10} is also included in all fits using a grid of
  XSTAR models \citep{walton13}. Fits to Observation 1 included a blackbody component to
  account for the soft excess below $\approx 0.6$~\kev. The blackbody
  temperature and normalization are consistent with those found by
  \citet{sam11}: $kT \approx 0.1$~keV and normalization $\approx
  10^{-4}$. The \fe\ line energy is fixed at 6.4~keV in these models. The
  seed photon temperature for the \texttt{compps} models is fixed at
  $8.9$~eV, and a \texttt{diskbb} spectrum is assumed. $\tau$ is
  estimated using the best fit $kT_e$ and $y$.}
\end{deluxetable}

To attempt to further characterize the \fe\ line and any associated
reflection continuum, the cutoff power-law is replaced with a \texttt{pexrav}
model \citep{mz95} where the abundances are fixed at solar and
assuming an inclination angle of $40\deg$ \citep{gio01}. However, no improvement is found with only an
upper-limit to the reflection fraction of $R < 0.03$ (the best-fit
value of $R$ is $10^{-15}$). Replacing \texttt{pexrav}
with other reflection models that include the \fe\ line, such as
\texttt{pexmon} \citep{nan07} or \texttt{relxill} \citep{gar14}, and
continuing to assume solar abundances, result in significantly worse values of
$\chi^2$ and $R \sim 0.1$--$0.2$. Allowing the iron abundance to fit
freely improves the fit ($\chi^2/$dof$=983/986$ for
\texttt{relxill}), but the iron abundance is constrained to be
$>8.6\times$ the Solar value, and the reflection fraction remains very
small ($R=0.07\pm 0.02$). The best fit ionization parameter from the
\texttt{relxill} model is $\log \xi=0.31_{-0.31}^{+1.41}$. This
produces a neutral reflection spectrum, consistent with the line
energy of $\approx 6.4$~keV found by the Gaussian fits. 

Relativistic effects can broaden the
\fe\ line if it originates from the inner accretion disk \citep[e.g.,][]{fab89,laor91,rn03}. In
this situation, the width of the line can be translated into a minimum
radius of the reflecting region along the accretion disk. To check if
relativistic effects are causing the observed line width in \threec\ the Gaussian \fe\ line in the
cutoff power-law model was replaced with a relativistic line profile
(using \texttt{relline}; \citealt{dau10}). The \texttt{relline} model
produces a good fit ($\chi^2/$dof$=981/988$) and gives an inner
radius of $82^{+94}_{-36}$~$r_g$ (where $r_g=GM/c^2$ is the
  gravitational radius of a black hole with mass $M$) for an
  inclination angle of $40\deg$ and an unbroken
  emissivity index of $-3$. This result indicates that the detected line likely does not
  originate from the inner accretion disk. Letting the inclination
  angle or emissivity profile vary does not improve the fit and yields
  only lower limits to the parameters (inclination angle$> 25$~deg;
  emissivity index$> 2.25$). No evidence for a weak relativistic
  line from closer to the inner disk is found in the data. The
  \texttt{relxill} model mentioned above also includes relativistic
  blurring and a similar constraint on the inner radius ($>57$~$r_g$)
  was found by that model.

The very weak reflection continuum found in this observation of
\threec\ presents an opportunity to determine if the continuum from a
broad-line radio galaxy is described by thermal Comptonization models
similarly to radio-quiet Seyfert galaxies. We therefore
remove the cutoff power-law component and replace it with a \texttt{compps}
model\footnote{Models using \texttt{comptt} \citep{tit94} provide similar fits, but
  with much larger uncertainties on the parameters.} \citep{ps96},
assuming no reflection component. The
seed photons are assumed to arise from a multi-temperature blackbody
(the \texttt{diskbb} model) with a maximum temperature
of $8.9$~eV, the characteristic temperature of an accretion disk around
a $10^9$~M$_{\odot}$ black hole \citep{mar04}, assuming accretion at
20\% of the Eddington limit \citep{glio07}. In addition to the plasma
temperature $kT_e$, the Compton $y$ parameter,
defined as $y=4\tau (kT_e/511~\mathrm{keV})$, is used as a fit
parameter in place of the optical depth $\tau$. The \texttt{compps} model predicts
spectra for several different geometries of the Comptonizing
plasma (e.g., slab, sphere, or cylinder). None of the geometries give
qualitatively different results, so we focus here on the fits from
the slab geometry with seed photons injected at the bottom of the slab, as it gives the
best constraints on the parameters, and it is one of the two
geometries that is accurately computed by \texttt{compps} (rather than
approximated; \citealt{ps96}). The spectral model
with \texttt{compps} and an \fe\ line provides an excellent fit to the data
($\chi^2/$dof$=976/988$) and gives the temperature of the
Comptonizing electrons to be $kT_e=330\pm 30$~\kev\ and a Compton $y$
parameter of $0.38^{+0.04}_{-0.05}$ (see Table~\ref{table:fits}). Together, these values result in
an optical depth of $\tau=0.15$. The $2$--$10$~\kev\
flux of \threec\ determined by this fit is $F_{2-10\
  \mathrm{keV}}=2.9\times 10^{-11}$~erg~cm$^{-2}$~s$^{-1}$, which
corresponds to an unabsorbed luminosity of $L_{2-10\ \mathrm{keV}} = 2.3\times
10^{44}$~erg~s$^{-1}$.

\subsection{Observation 1 --- High Flux}
\label{sub:obs2}
The analysis of the short Observation~1 is significantly improved with
the inclusion of the \swift-XRT data. The left-hand panel of
Fig.~\ref{fig:data} plots the \swift\ and \nustar\ count rate spectra,
as well as the residuals to a power-law fit (ignoring the
$4$--$7.5$~\kev\ band). The fit is adequate ($\chi^2/$dof$=810/791$),
but there are clear positive residuals at $\la 0.7$~\kev, and hints of an
\fe\ line. The soft excess has been seen in earlier observations of
\threec\ \citep[e.g.,][]{sam11}, and is well modeled by a blackbody component. The
addition of the blackbody significantly improves the fit (new
$\chi^2/$dof$=796/789$; F-test probability $> 99.9$\%) and gives a
temperature of $kT=0.09^{+0.03}_{-0.02}$~\kev\ and
normalization\footnote{Equal to $L_{39}/D_{10}^2$ where $L_{39}$ is
  the luminosity of the blackbody in units of $10^{39}$~erg~s$^{-1}$,
  and $D_{10}$ is the distance to \threec\ in units of 10~kpc.}
$\approx 1.3\times 10^{-4}$, consistent with previous measurements
\citep{sam11}. These values change very little for the different models of the
high energy emission. The addition of a high-energy cutoff makes only
a marginal improvement to the fit ($\chi^2/$dof$=793/788$; F-test
probability $\approx 90$\%) with $E_{\mathrm{cut}} > 167$~\kev. A
narrow ($\sigma=90$~eV) \fe\ line at 6.4~\kev\ results in
$\chi^2/$dof$=784/787$ which is significant at the $99.7$\% level
according to the F-test. Allowing $\sigma$ to vary reduces $\chi^2$ to
$780$ for $786$~dof ($96$\% significant) and gives
$\sigma=0.37^{+0.3}_{-0.2}$~\kev. The lower-limit to the cutoff energy
now rises to $190$~\kev. The flux of the line is $\approx 30$\%
larger than in Observation 2 (see Table~\ref{table:fits}), but is consistent with a
constant flux across both observations (Fig.~\ref{fig:linecontours}). Although the high-energy cutoff is not
strictly necessary for this fit, we show the parameters of this model
in the upper-half of Table~\ref{table:fits} to compare with the results from
Observation 2. As with Observation 2, \texttt{pexrav} only gives an
upper limit to the reflection fraction $R < 0.16$ (the best fit value
of $R$ is $0.03$, but is consistent with $0$ within the errors). The
\texttt{pexmon} and \texttt{relxill} models both give good fits
($\chi^2/$dof$=787/787$ and $786/785$, respectively) with solar
abundances, but with low reflection fractions ($0.13\pm 0.09$ for the
\texttt{pexmon} model and $0.10^{+0.07}_{-0.06}$ for the
\texttt{relxill} model). The \texttt{relxill} model yields an
upper limit to the ionization parameter of $\log \xi < 3.3$, and is
unable to constrain an inner radius from disk reflection. Given the
poor statistics of this short observation, no further modeling of the
\fe\ line was attempted.

The combination of \nustar\ and \swift-XRT produces a spectrum that
spans over 2 orders of magnitude in energy which, in combination with
the weak reflection continuum, provides a significant lever arm for
potentially constraining thermal Comptonization models (even with the
weak constraint on $E_{\mathrm{cut}}$). Applying the same \texttt{compps}
model as in Observation 2 yields an excellent fit to Observation 1
($\chi^2/$dof$=778/786$) with $kT_e=231^{+50}_{-88}$~\kev\ and
$y=0.40^{+0.07}_{-0.06}$, corresponding to $\tau=0.23$ (Table~\ref{table:fits}). The
$2$--$10$~\kev\ flux and unabsorbed luminosity obtained by this model is
$5\times 10^{-11}$~erg~cm$^{-2}$~s$^{-1}$ and $4\times
10^{44}$~erg~s$^{-1}$, respectively. Thus, \threec\ was $1.7\times$
fainter in Observation 2 than in Observation 1.

\section{Discussion and Conclusions}
\label{sect:discuss}
\subsection{Comparison to Previous Results}
\label{sub:prior}
The results from the \nustar\ observations agree with many of the
previous measurements of \threec\ \citep[e.g.,][]{grandi01}. In particular, the long \suzaku\ observation
analyzed by \citet{sam11} and \citet{walton13} is the highest quality
spectrum taken of \threec\ and these authors find a similarly small
value of the reflection fraction ($R\approx
0.1$--$0.15$). \citet{sam11} combine the \suzaku\ data with the
integrated \swift-BAT spectrum to measure a cutoff energy of
$E_{\mathrm{cut}} = 175^{+25}_{-20}$~\kev, consistent with both
\nustar\ measurements (Table~\ref{table:fits}). The photon index and flux
obtained from the \suzaku\ data were $\Gamma=1.74$ and
$F_{2-10\ \mathrm{keV}}=4.1\times 10^{-11}$~erg~cm$^{-2}$~s$^{-1}$
\citep{sam11}. These values nicely fall in the middle of the two
\nustar\ observations and support the assertion that \threec\ follows
the typical Seyfert pattern of softening as it brightens. 

The \fe\ line flux is measured to be $\approx 4\times
10^{-5}$~ph~cm$^{-2}$~s$^{-1}$ in both observations despite a change
in continuum flux by nearly a factor of two. Previous observations with
instruments with similar energy resolutions at 6.4~\kev\ all find line
fluxes at approximately this value
\citep{esm00,grandi01,glio07}. Indeed, \citet{grandi01} noted that
line flux has remained constant in all earlier observations despite
large changes in the continuum flux, and argued that the line must
result from a distant reprocessor. The \nustar\ observations are
consistent with this interpretation; however, this conclusion does not
exclude the presence of a weak, relativistically broadened component
to the line as described by \citet{sam11}. To test this possibility, we attempted
to model Observation 2 with one broad and one narrow Gaussian to model the \fe\ line
complex. While a good fit is achieved, the $\chi^2$ is unchanged from
the single Gaussian model, and the normalization of the narrow
6.4~\kev\ line is consistent with zero. A similar result is achieved
when attempting a double reflection model using two \texttt{relxill}
models; namely, the normalization of the second reflector was sent
towards zero. Again, this is not evidence that this model is
incorrect, but our data around the \fe\ line does not allow us to test
these complex models.

\subsection{A Comptonizing Corona}
The spectral analysis of the \nustar\ data presents evidence that the X-ray spectrum
of the BLRG \threec\ exhibits a Seyfert-like Comptonizing
corona; specifically, the corona cools and produces a softer spectrum
when it brightens (Table~\ref{table:fits}). This behavior is inferred
to be typical for non-jetted Seyfert~1 galaxies \citep[e.g.,][]{lub10,vvp11}, but is not
observed in AGNs with jet-dominated continua (e.g., 3C~273;
\citealt{cher07})\footnote{Interestingly, some X-ray binaries also do not
  show the `softer-when-brighter' behavior when in the jet-dominated
  state \citep[e.g.][]{gan08}.}. The X-ray spectra of jet-dominated AGNs (e.g.,
blazars) typically have
$\Gamma < 1.5$ \citep[e.g.,][]{sam06,gia11}, produce hard X-ray
variations $\approx 2\times$ larger than Seyferts \citep[e.g.,][]{cher07,soldi14},
and are gamma-ray sources rather than exhibiting cutoffs at $\sim
100$--$200$~\kev\ \citep{gio06}. In contrast, \threec\ is not a \textit{Fermi}
source \citep{nol12}, has hard X-ray variability properties similar to
Seyfert~1s \citep{soldi14},
and has a detectable X-ray spectral cutoff. The hard X-ray properties
of \threec\ revealed by \nustar\ are therefore entirely consistent with
those found from non-jetted Seyfert~1s.

\begin{figure*}
\includegraphics[width=0.85\textwidth]{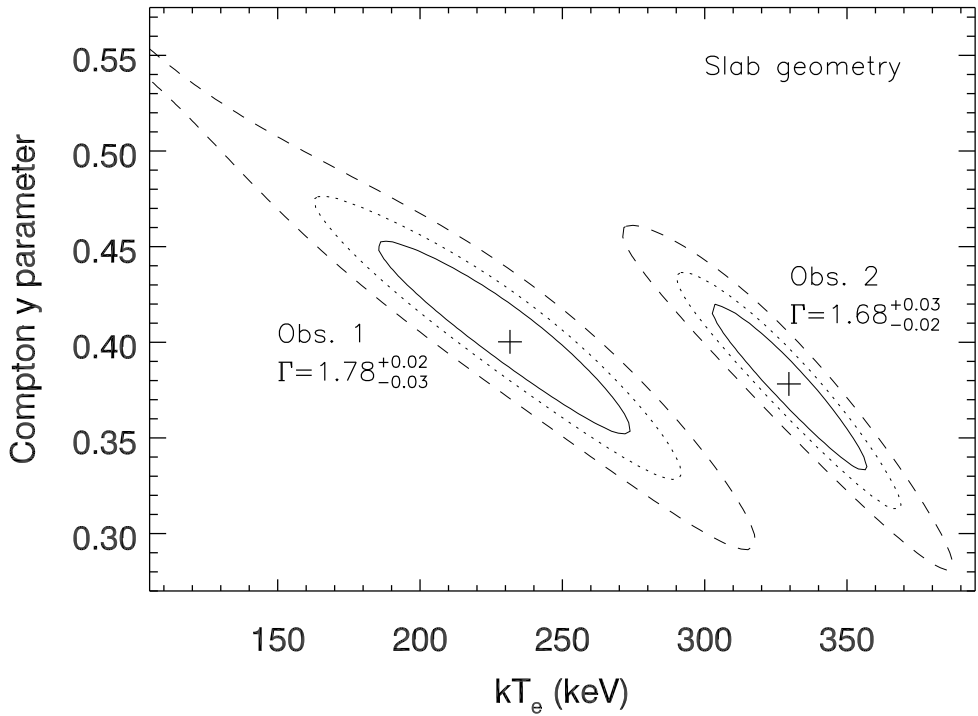}
\caption{The 68\% (solid), 90\% (dotted) and 95\% (dashed) joint confidence
  contours on $kT_e$ and $y$ calculated for Observations 1 and 2
  from the best fitting \texttt{compps} + 6.4~\kev\ Gaussian models
  (Table~\ref{table:fits}). The crosses mark the location of the best
  fit values.}
\label{fig:contour}
\end{figure*}

To examine the changes inferred in the corona properties of \threec\ more closely, confidence contours for
$kT_e$ and $y$ are computed using the best fitting \texttt{compps} model and
are overplotted in Figure~\ref{fig:contour}.
The contours show that the change in coronal temperature between the
two observations is $\approx 90$\% significant, and that the corona
is cooler in the more luminous state \citep[e.g.,][]{hmg97,grandi01}. This change in temperature is
consistent with the model of a thermal Comptonizing corona producing the X-ray spectrum in
\threec. Interestingly, the value of $kT_e$ derived from the \texttt{compps} model of
Observation 2 is larger than the common approximation $kT_e \approx
E_{\mathrm{cut}}/2$ (for $\tau \la 1$) or $kT_e \approx
E_{\mathrm{cut}}/3$ (for $\tau \gg 1$)\footnote{Observation 2 can be
  well fit with a low-$kT_e$, high-$\tau$ model (e.g., $kT_e \approx
  41$~keV and $\tau \approx 2.5$ assuming spherical geometry), but with
  a larger $\chi^2$ than the high-$kT_e$ fit discussed here ($\Delta \chi^2 \approx
  +11)$). We therefore focus on the latter model in this paper.}. This mismatch has been seen before by \citet{pet01} and is simply
explained by recalling that cutoff power-laws are only approximations
to the spectra produced by thermal Comptonization spectra. Geometric
effects due to the anisotropic nature of the Comptonization process
will introduce a slight curvature to the spectrum at high
energies \citep{stern95,ps96}. The effect is especially important for corona with high
$kT_e$ and low $\tau$ \citep{pet00,pet01}. Interestingly, the
\texttt{compps} fits yield lower $\chi^2$ than the cutoff power-law model for
both \threec\ datasets indicating that this curvature may indeed be in
the data. Moreover, although $kT_e$ and $\tau$ remain similar,
\texttt{compps} models where the seed photons are distributed
isotropically in the hot plasma result in $\Delta \chi^2=+5$ compared to the
anisotropic model reported in Table~\ref{table:fits}. This result provides
tantalizing evidence that anisotropic Comptonization is playing an
important role in \threec, and strongly supports the use of
sophisticated Comptonization models such as \texttt{compps} when
interpreting the high-energy cutoffs of AGNs. In particular,
Comptonization modeling of AGNs with low $R$ may prove especially
useful in determining details of the corona.

\subsection{Comparison to Other Coronal Measurements}
Currently, \nustar\ has measured the coronal parameters in two other
AGNs: the Seyfert 1 galaxy IC~4329A ($kT_e =61\pm 1$~keV and
$\tau=0.68\pm 0.02$; \citealt{bren14b}) and the narrow-line Seyfert 1
Swift~J2127.4+5654 ($kT_e=68^{+37}_{-32}$~keV and
$\tau=0.35^{+0.35}_{-0.19}$; \citealt{mar14}). Both these measurements
assume a slab geometry and were performed using the \texttt{comptt}
model. \threec\ with a $kT_{e} \ga 150$~keV appears to have a
significantly hotter corona than either of these sources, and,
potentially, a more tenuous corona than IC~4329A. However,
\citet{matt14} also report a high temperature corona ($kT_e \approx
110$--$210$~keV) in joint \xmm-\nustar\ fits of the Seyfert~1
Ark~120. Thus, at this early stage, we cannot draw a conclusion about
the importance of the hot corona observed in \threec, but $\approx 20$ bright
AGNs have been targeted by \nustar\ with the goal of determining the
coronal parameters \citep{harr13}. This sample will allow interesting
comparisons of coronal temperatures and optical depths among AGNs of
different classes.

\subsection{Implications on Coronal Geometry and Dynamics}
\label{sub:coronal}
\begin{figure}
\includegraphics[width=0.375\textwidth,angle=-90]{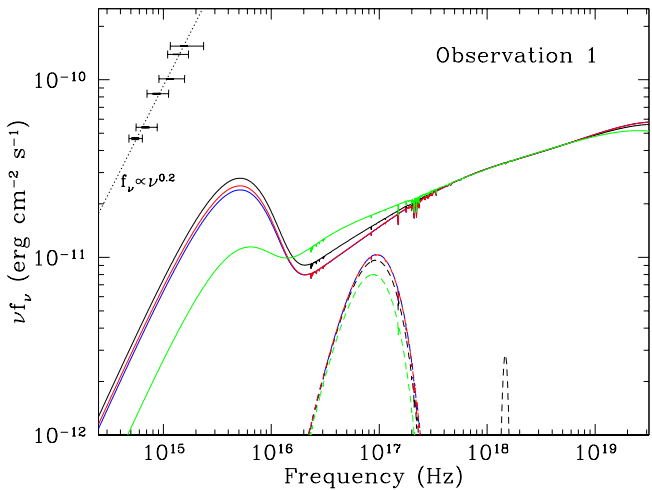}
\caption{Predicted optical/UV/X-ray SED for Observation 1 of
  \threec\ as determined by the \texttt{compps} modeling of the X-ray
  data. Galactic absorption has been removed from the models, but the
  small effects of
  the warm absorber are still visible. To highlight the Comptonization
  spectra, the full SED has been decomposed into the \texttt{compps}
  continuum (solid lines), the soft excess and the \fe\ line (dashed
  lines). The various lines show how the predictions depend on different coronal
  geometries: slab (black), cylinder (blue), hemisphere (red), and
  sphere (green). The same seed photon distribution
  (\texttt{diskbb} with $kT=8.9$~eV) was used for all the
  Comptonization models. The data points are from the
  \swift\ UVOT observations during Observation 1, and have been corrected for
  Galactic and internal extinction (see Sect.~\ref{sect:obs}).}
\label{fig:sed}
\end{figure}

Recent observational innovations utilizing \fe\ reverberation and
microlensing have allowed the first constraints to be placed on the
size of the X-ray corona in radio-quiet AGNs \citep[e.g.,][]{chartas09,zog12,wf13}. These techniques
all point to a small, centrally concentrated corona, situated within
$\approx 20$~$r_g$ from the black hole \citep{rm13}. Such compact coronae
provide the necessary illumination of the inner disk to produce a relativistic
\fe\ line \citep{fab14}. Although the \threec\ spectra analyzed here do not
show strong evidence for a relativistic line, we can combine the
Comptonization modeling and the simultaneous \swift\ data during
Observation 1 to obtain a simple view of the X-ray corona of
\threec. Figure~\ref{fig:sed} shows the optical/UV/X-ray SED of \threec\ predicted by the best fitting
\texttt{compps} models of Observation 1. The different lines show
how the predicted SED vary for different coronal geometries. All of
these models assume the same seed photon spectrum (a \texttt{diskbb}
with maximum $kT=8.9$~eV) and give the same $\chi^2$ to the X-ray
data.
The data points in the figure are the de-reddened fluxes of
\threec\ obtained from the simultaneous \swift\ observation (see
Sect.~\ref{sect:obs}). A least-squares fit to the \swift-UVOT data results
in $f_{\nu} \propto \nu^{0.2\pm 0.1}$ (dotted line in
Fig.~\ref{fig:sed}), consistent with the 
$f_{\nu} \propto \nu^{0.3}$ slope expected for a Shakura-Sunyaev
accretion disk \citep[e.g.,][]{fkr02}. Interestingly, the amplitude of the
\swift-UVOT data is $\sim 10\times$ larger than the \texttt{compps}
models. The UV emission predicted by the Comptonization models is
from the zeroth-scattering order, which, in a low $\tau$ corona such
as seen in \threec, is nearly equal to the total UV emission necessary to
produce the observed X-rays. Thus, the
SED models indicate that the X-ray corona may intercept only $\sim 10$\%
of the accretion disk UV flux, broadly consistent with
models of patchy corona in Seyfert galaxies \citep[e.g.,][]{hmg94}. 

Evidence for a Comptonizing corona in \threec, when combined with the
fact that the high-energy variability properties of BLRGs are
consistent with those from non-jetted Seyfert 1s \citep{soldi14}, shows that
contamination from jets in the X-ray band of BLRGs has likely been
minimal. Thus, there must be another explanation for the extremely weak reflection
features of \threec. Interestingly, the source does present a significant (EW $\approx
100$~eV) \fe\ line, but it does not originate from the inner accretion
disk, nor does it arise from the broad line region (the \fe\ FWHM$\sim
30,000$~km~s$^{-1}$, compared to $\approx 12,000$~km~s$^{-1}$ for the H$\alpha$
line; \citealt{eh94}). However, the line width is only slightly larger
than the energy resolution ($400$~eV FWHM; \citealt{harr13}), so it is
possible that the observed \fe\ line width and strength is caused by
the blending of
multiple line components. Regardless of the number of components to
the line, there does not seem to be a significant reflection continuum
in the observed spectrum. This fact points to an origin in material
with a large Fe abundance or from Compton thin clouds that
may originate in a dusty outflow from the nucleus
\citep[e.g.,][]{honig13}.

If the X-ray source in \threec\ is a Comptonizing corona similar to
Seyfert galaxies, then the lack of reflection from the inner disk is
still puzzling. The inner accretion disk must be either blocked from
view, absent, or emits such a weak reflection
continuum that it cannot be identified in our data. \citet{sam11} present
evidence for a highly ionized inner accretion disk in \threec\ that, when combined with a
distant reflector that fits the narrow component of the \fe\ line,
also accounts for the soft excess. Unfortunately, our data quality is
not high enough to directly test this model, although we find a
slightly broader neutral \fe\ line than \citet{sam11} in both
observations. Another possibility for the extremely weak disk reflection is a
transition to an optically thin, radiatively inefficient accretion flow
inside $\sim 70$--$100$~$r_g$ as a result of a `jet cycle' where the
inner disk is `emptied' during a jet outburst \citep[e.g.,][]{loh13}. However, there
was no radio monitoring of \threec\ during the \nustar\ observations,
so this model cannot be tested for this source. Finally, if the X-ray
emitting corona is outflowing away from the accretion disk \citep{belo99,mbp01} then
it would naturally explain the hard power-law slope and the apparently
very weak disk reflection (see also \citealt{fab14}). The models of \citet{mbp01} indicate that an outflow
speed of $\ga 0.5c$ would yield $\Gamma \sim 1.7$ and $R \la
0.2$. Given that the black hole environment of \threec\ is producing a
relativistic jet, this process, combined with ionized reflection from
close to the black hole, may
be a promising model to explain the weak reflection features seen in
BLRGs. Interestingly, IC 4329A also has weak reflection features and a
hard continuum similar to \threec\ \citep{bren14b}, but does not produce a large
scale radio-jet. \citet{bren14b} estimate an outflow speed of $\sim
0.2c$ for that source. These results may be hinting at a deeper
connection between the corona and the base of a jet
\citep[e.g.,][]{markoff05}. Further discrimination of the models will require high
resolution and broadband X-ray data (with, e.g.,
\textit{ASTRO-H}) combined with contemporaneous radio imaging.
 
\acknowledgments
We thank the referee for a helpful report that improved the paper.
This work was supported under NASA Contract No.\ NNG08FD60C, and made
use of data from the \nustar\ mission, a project led by the California
Institute of Technology, managed by the Jet Propulsion Laboratory, and
funded by the National Aeronautics and Space Administration. We thank
the \nustar\ Operations, Software and Calibration teams for support
with the execution and analysis of these observations. This research
has made use of the \nustar\ Data Analysis Software (NuSTARDAS)
jointly developed by the ASI Science Data Center (ASDC, Italy) and the
California Institute of Technology (USA). This work made use of data supplied by the UK Swift Science Data Centre at the
University of Leicester. This research has made use of data, software
and/or web tools obtained from NASA's High Energy Astrophysics Science
Archive Research Center (HEASARC), a service of Goddard Space Flight
Center and the Smithsonian Astrophysical Observatory. DRB acknowledges support
from NASA ADAP grant NNX13AI47G and NSF award AST 1008067. AM acknowledge financial support from Italian Space Agency under grant ASI/INAF I/037/12/0-011/13 and 
from the European Union Seventh Framework Programme (FP7/2007-2013)
under grant agreement n.312789. M.\,B.
acknowledges support from the International Fulbright Science and
Technology Award.

{}

\end{document}